\documentstyle[amssymb,twocolumn,aps,pra,psfig]{revtex}

\begin{document}
\title{Quantum limits of cold damping with optomechanical coupling}
\author{Jean-Michel Courty\thanks{%
courty@spectro.jussieu.fr}, Antoine Heidmann\thanks{%
heidmann@spectro.jussieu.fr} and Michel Pinard\thanks{%
pinard@spectro.jussieu.fr}}
\address{Laboratoire Kastler Brossel\thanks{%
Unit\'{e} mixte de recherche de l'Universit\'{e} Pierre et Marie Curie, de
l'Ecole Normale Sup\'{e}rieure associ\'{e} et du Centre National de la
Recherche Scientifique}\thanks{%
website: www.spectro.jussieu.fr/Mesure}, Case 74, 4 place Jussieu, 75252\\
Paris Cedex 05, France}
\date{October 12, 2001}
\maketitle

\begin{abstract}
Thermal noise of a mirror can be reduced by cold damping. The displacement
is measured with a high-finesse cavity and controlled with the radiation
pressure of a modulated light beam. We establish the general quantum limits
of noise in cold damping mechanisms and we show that the optomechanical
system allows to reach these limits. Displacement noise can be arbitrarily
reduced in a narrow frequency band. In a wide-band analysis we show that
thermal fluctuations are reduced as with classical damping whereas quantum
zero-point fluctuations are left unchanged. The only limit of cold damping
is then due to zero-point energy of the mirror.
\end{abstract}


\section{Introduction}

Characterization and control of thermal noise is of particular interest for
very sensitive measurements such as interferometric gravitational-wave
detectors\cite{Bradaschia90,Abramovici92}.\ Fluctuations of the mirror
position result from the thermal excitation of various mechanical modes of
the suspended mirrors, corresponding either to external degrees of freedom
of the suspension system or to acoustic modes of the mirror substrate. This
leads to undesirable displacements of the mirrors and limits the sensitivity
of the measurement. For example internal thermal noise is due to
deformations of the mirror surface and constitutes the main limitation of
gravitational-wave detectors in the intermediate frequency domain.

Thermal fluctuations are associated with dissipation mechanisms inherent in
the system\cite{Einstein05,Nyquist28,LandauCh8} and are therefore very
difficult to avoid. Apart from passive methods such as the modification of
mechanical damping\cite{Rowan00} or cryogenic methods to lower the
temperature\cite{Uchimaya98}, these thermal fluctuations may be reduced by
using active systems. In particular it has been proposed to use an
optomechanical displacement sensor to monitor the brownian motion of a
mirror \cite{Mancini98,Cohadon99}. Cold damping techniques have also been
studied to reduce the effective temperature of a system well below the
operating temperature\cite{Milatz53,Grassia00}. They have been proposed to
reduce the brownian motion of an electrometer\cite{Bernard91} and used to
achieve very high sensitivity in accelerometers developped for fundamental
physics applications in space\cite{Touboul92}.

Such techniques have been successfully applied to an optomechanical system
composed of a high-finesse cavity and a feedback loop\cite
{Cohadon99,Pinard00}. The displacement of the mirror is measured by the
optical Fabry-Perot cavity with a very high sensitivity\cite
{Hadjar99,Tittonen99}. This information is fed back to the mirror via the
radiation pressure of an intensity-modulated laser beam. For an appropriate
design of the feedback loop, the radiation pressure exerted on the mirror is
proportional to the mirror velocity. The servo-control force then
corresponds to a viscous force. In contrast to passive damping which is
necessarily accompanied by thermodynamic fluctuations\cite{LandauCh8}, this
active damping does not add any thermal noise and allows to greatly reduce
the brownian motion of the mirror. Power noise reductions around the
mechanical resonance frequency of the mirror as large as 1000 have been
experimentally obtained. As proposed in \cite{Cohadon99} such short cavities
could in principle be used to monitor the mirrors of a gravitational-wave
interferometer.\ They would be insensitive to a gravitational wave and allow
to reduce the mirrors thermal noise without affecting the response of the
interferometer to the gravitational signal.

In the experimental conditions of references\cite{Cohadon99,Pinard00},
quantum effects are negligible as compared to thermal noise so that a
classical treatment of the system is satisfactory to fit the experimental
results. The analysis of electromechanical devices has provided precise
discussions of the classical limits of cold damping\cite{McCombie53}. It has
in particular been pointed out that thermal fluctuations of position can be
reduced by electronic damping of the motion without classical limit\cite
{Ritter85}.

For an analysis of the actual limits of cold damping techniques, it is
however essential to consider quantum fluctuations\cite{Grassia00}. It is
well known that quantum fluctuations play a fundamental role in the limits
of sensitivity for interferometric position measurements. The respective
role of the phase noise of the detection beam and of its intensity noise
through radiation pressure effects on the mirrors has been thoroughly
analysed. This has led to the definition of a standard quantum limit \cite
{Caves81,Braginsky92} and of an ultimate quantum limit \cite{Jaekel90}.
Reduction of quantum noise of light has also motivated studies on the
behaviour of quantum noise in presence of active feedback\cite
{Haus86,Wiseman94}. It has been shown that this technique allows to
eliminate back action in quantum measurements\cite{Wiseman95} or to reduce
quantum fluctuations of a light beam inside a feedback loop below the
standard quantum limit, allowing an increased sensitivity of measurements
\cite{Buchler99}.

With the optomechanical system, it is possible to reduce the initial
temperature of the mirror and to improve the experimental efficiency of the
cooling.\ This opens the way to a quantum regime of cold damping and raises
several questions. For passive systems, in the limit of a null temperature,
thermodynamic fluctuations associated with dissipation reproduce the quantum
fluctuations required by Heisenberg inequalities\cite{Callen51}. What
happens to the fluctuations of an actively cooled system? Does cold damping
allow to reduce quantum fluctuations as it does with classical fluctuations?
Is there a temperature limit for this technique? Do the quantum fluctuations
of light introduce a limit in the efficiency of the cooling process? How is
it possible to reach the limits?

In this paper, we address these questions by a quantum analysis of cold
damping with optomechanical coupling.\ We use a theoretical treatment based
on quantum networks theory\cite{Yurke84,Gardiner88}. This approach allows to
treat in the same framework both thermal and quantum fluctuations, for
passive as well as for active elements\cite{Grassia00,Grassia00b}. This
quantum description ensure the consistency of the approach and allows to
study the effect of cold damping for very low initial temperature. It has
for example been proven fruitful in the analysis of quantum limits for
ultrasensitive measurements with cold damped capacitive accelerometers\cite
{Grassia00}.

We show that the only limit of cold damping are zero-point quantum
mechanical fluctuations of the mirror, in agreement with general
thermodynamical relations and Heisenberg inequalities\cite{Callen51}. This
limit originates in our system from quantum fluctuations of the light beam
used in the displacement measurement. As in the case of interferometric
measurements, quantum noise of light limits the sensitivity of the
measurement by the high-finesse cavity, and quantum fluctuations of
radiation pressure disturb the mirror motion. In contrast with
interferometers, however, radiation pressure effects are controlled by the
active feedback.\ It is then possible to reduce the mirror fluctuations down
to the zero-point quantum mechanical fluctuations, by an appropriate
optimization of the cold damping mechanism.

In section\ \ref{M_OptomechanicalTransducer} we present the optomechanical
transducer composed of the high-finesse cavity and the feedback loop.\ Basic
relations of cold damping mechanism are derived in section \ref
{M_ColdDamping}.

In section \ref{M_Resonance} we analyze the quantum limits for the reduction
of mirror thermal noise on a narrow frequency band around the mechanical
resonance. We show that it is possible to reduce this noise to arbitrary low
values.

In section \ref{M_Temperature} we consider the whole noise spectrum of the
mirror motion. We show that the temperature may be reduced to zero with a
noise spectrum corresponding to zero-point fluctuations of the mirror.

We finally establish in section \ref{M_QuantumLimit} the general limits of
noise in cold damping mechanisms. We show that the optomechanical system may
provide a mean to reach these limits. The effects of this active technique
are then equivalent to a coupling with a thermodynamic reservoir at a null
temperature.

\section{Optomechanical transducer}

\label{M_OptomechanicalTransducer}
\begin{figure}
\centerline{\psfig{figure=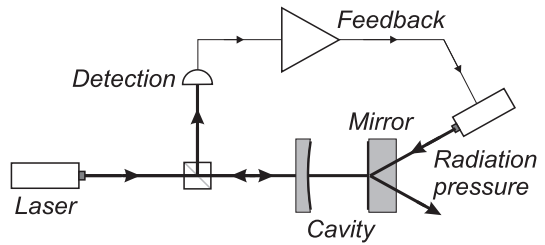,width=7cm}}
\vspace{2mm}
\caption{Scheme of the system studied in the paper. An optomechanical transducer 
made of a high-finesse cavity is used to monitor the thermal noise of a mirror. This signal
is fed back on the mirror via the radiation pressure of an intensity-modulated laser beam}
\label{Fig_Setup}
\end{figure}

The general scheme of the system is shown
in figure \ref{Fig_Setup}. It is based on a measurement of the thermal noise
of the mirror and on a feedback loop which applies a properly adjusted force
on the mirror. The mirror is the end mirror of an optomechanical transducer
made of a single-port high-finesse cavity resonant with the incident laser
beam.\ Detection of the phase of the reflected field provides a signal
proportional to the mirror displacement.\ This signal is used to implement
the feedback loop via the radiation pressure of an intensity-modulated laser
beam reflected from the back of the mirror. This modulated beam exerts a
force proportional to the mirror displacement with a transfer function that
can be tailored by the electronics. In order to apply a viscous damping
force, the transfer function has to be adjusted in such a way that the force
is proportional to the mirror velocity. Practical implementation can consist
in an electronic derivation of the signal delivered by the optomechanical
transducer\cite{Cohadon99}.

The electromagnetic field in the high-finesse cavity is described by a
single harmonic mode characterized by annihilation and creation operators $a$
and $a^{\dagger }$. To determine the input-output relations for the
fluctuations we linearize the evolution equations around the steady state of
the system\cite{Reynaud89}. For a resonant cavity the fluctuations $a^{{\rm %
in}}\left[ \Omega \right] $, $a\left[ \Omega \right] $ and $a^{{\rm out}}%
\left[ \Omega \right] $ for the complex amplitudes at frequency $\Omega $ of
the incident, intracavity and reflected fields are related by 
\begin{mathletters}
\label{Eq_A}
\begin{eqnarray}
-i\Omega \tau a &=&-\gamma a+\sqrt{2\gamma }a^{{\rm in}}+i\varkappa X,
\label{Eq_Aa} \\
a^{{\rm out}} &=&-a^{{\rm in}}+\sqrt{2\gamma }a,
\end{eqnarray}
where the Fourier transform $a\left[ \Omega \right] $ of the time-dependent
operator $a\left( t\right) $ is defined as 
\end{mathletters}
\begin{equation}
a\left[ \Omega \right] =\int a\left( t\right) e^{i\Omega t}dt.
\end{equation}
Equation (\ref{Eq_Aa}) determines the dynamics of the intracavity field.\ $%
\tau $ is the round trip time of the cavity, $\gamma $ is its damping rate
(for a lossless cavity, $1-\gamma $ and $\sqrt{2\gamma }$ are respectively
the reflection and transmission of the input mirror, with $\gamma \ll 1$).
The second equation is the input-output relation for the fields.\ These
equations are the usual ones for a single-ended cavity, with an extra term
in eq.\ (\ref{Eq_Aa}) which couples the intracavity field to the mirror
position $X$.\ Neglecting any retardation effect\cite{Tourrenc85} a
displacement of the mirror induces a phase shift for the intracavity field
proportional to the change of the optical path followed by the light beam.\
The optomechanical coefficient $\varkappa $ is given by\cite{Pinard99}

\begin{equation}
\varkappa =2k_{0}\alpha _{0},  \label{Eq_K}
\end{equation}
where $k_{0}$ is the field wavevector and $\alpha _{0}$ is the mean
intracavity field.

The input field operators $a^{{\rm in}}$ and $a^{{\rm in}\dagger }$ obey the
free fields commutation relations

\begin{mathletters}
\label{Eq_Dummy0}
\begin{eqnarray}
\left[ a^{{\rm in}}\left[ \Omega \right] ,a^{{\rm in}\dagger }\left[ \Omega
^{\prime }\right] \right] &=&2\pi \delta \left( \Omega +\Omega ^{\prime
}\right) , \\
\left[ a^{{\rm in}}\left[ \Omega \right] ,a^{{\rm in}}\left[ \Omega ^{\prime
}\right] \right] &=&\left[ a^{{\rm in}\dagger }\left[ \Omega \right] ,a^{%
{\rm in}\dagger }\left[ \Omega ^{\prime }\right] \right] =0.
\end{eqnarray}
Since a resonant cavity does not introduce any phase shift between the
incident, intracavity, and output mean fields, the complex amplitudes of
these fields can be simultaneously taken real.\ Amplitude and phase
quadratures $a_{1}$ and $a_{2}$ of the field then correspond to the real and
imaginary parts of the operator $a$, 
\end{mathletters}
\begin{mathletters}
\label{Eq_Dummy1}
\begin{eqnarray}
a_{1}\left[ \Omega \right] &=&a\left[ \Omega \right] +a^{\dagger }\left[
\Omega \right] , \\
a_{2}\left[ \Omega \right] &=&-i\left( a\left[ \Omega \right] -a^{\dagger }%
\left[ \Omega \right] \right) .
\end{eqnarray}
Assuming the incident field to be in a coherent state, quantum fluctuations
of the two input quadratures $a_{1}^{{\rm in}}$ and $a_{2}^{{\rm in}}$ are
characterized by 
\end{mathletters}
\begin{mathletters}
\label{Eq_Sain}
\begin{eqnarray}
\sigma _{a_{1}a_{1}}^{{\rm in}}\left[ \Omega \right] &=&\sigma
_{a_{2}a_{2}}^{{\rm in}}\left[ \Omega \right] =1, \\
\sigma _{a_{1}a_{2}}^{{\rm in}}\left[ \Omega \right] &=&0,
\end{eqnarray}
where the correlation functions $\sigma _{a_{i}a_{j}}^{{\rm in}}$ are
defined from the quantum average of the symmetrized product of operators $%
a_{i}^{{\rm in}}$ and $a_{j}^{{\rm in}}$, 
\end{mathletters}
\begin{equation}
\left\langle a_{i}^{{\rm in}}\left[ \Omega \right] \cdot a_{j}^{{\rm in}}%
\left[ \Omega ^{\prime }\right] \right\rangle =2\pi \delta \left( \Omega
+\Omega ^{\prime }\right) \sigma _{a_{i}a_{j}}^{{\rm in}}\left[ \Omega %
\right] .
\end{equation}

We now give the fundamental equations for the mirror motion. We assume that
the mechanical properties of the mirror can be described as a single
harmonic oscillator. Experimentally, mirror motion may result from the
excitation of many internal and external acoustic modes.\ The description as
a single oscillator is however a good approximation when frequencies are
limited to a small bandwidth around one mechanical resonance, by using for
example a bandpass filter either in the detection or in the feedback loop 
\cite{Cohadon99}.

We describe the mirror motion by the Fourier transform at frequency $\Omega $
of the mirror velocity, 
\begin{equation}
V\left[ \Omega \right] =-i\Omega X\left[ \Omega \right] .
\end{equation}
In the framework of linear response theory\cite{LandauCh12}, the velocity
linearly depends on applied forces which correspond to an external force $F_{%
{\rm ext}}$, a fluctuating force associated with damping and describing the
coupling with a thermal bath, and the radiation pressure of the intracavity
field, 
\begin{equation}
Z_{{\rm m}}V=F_{{\rm ext}}-\sqrt{2\hbar \left| \Omega \right| H_{{\rm m}}}m^{%
{\rm in}}+\hbar \varkappa a_{1},  \label{Eq_V}
\end{equation}
where $Z_{{\rm m}}$ is the mechanical impedance of the mirror. For a
harmonic oscillator of mass $M$, resonance frequency $\Omega _{{\rm m}}$ and
mechanical damping $H_{{\rm m}}$, this impedance has the simple form, 
\begin{equation}
Z_{{\rm m}}=M\left( -i\Omega +\frac{\Omega _{{\rm m}}^{2}}{-i\Omega }\right)
+H_{{\rm m}}.
\end{equation}
Note that we have assumed for simplicity that the mechanical oscillator is
viscously damped, that is $H_{{\rm m}}$ is independent of frequency. This
damping coefficient is related to the mechanical quality factor $Q$ by 
\begin{equation}
Q=\frac{M\Omega _{{\rm m}}}{H_{{\rm m}}}.  \label{Eq_Q}
\end{equation}

Last term in eq.\ (\ref{Eq_V}) represents the optomechanical coupling
between the mirror and the intensity of light in the cavity. It corresponds
to quantum fluctuations of radiation pressure\cite{Meystre83}. In the
linearized approach this non-linear coupling reduces to a term proportional
to the amplitude quadrature $a_{1}$ of the intracavity field \cite{Pinard99}%
.\ The optomechanical coefficient $\varkappa $ is the same as in eq.\ (\ref
{Eq_Aa}).

Quantum and thermal fluctuations associated with the damping can be deduced
from fluctuation-dissipation theorem\cite{Callen51} and appear in eq.\ (\ref
{Eq_V}) as an additional term proportional to an input field $m^{{\rm in}}$.
This field is the quantum analog to the usual Langevin force associated with
thermal fluctuations for a damped mechanical oscillator coupled to a thermal
bath at high temperature.\ It obeys the following commutation relation,

\begin{equation}
\left[ m^{{\rm in}}\left[ \Omega \right] ,m^{{\rm in}}\left[ \Omega ^{\prime
}\right] \right] =2\pi \delta \left( \Omega +\Omega ^{\prime }\right)
\varepsilon \left( \Omega \right) ,  \label{Eq_Commut}
\end{equation}
where $\varepsilon \left( \Omega \right) $ denotes the sign of the frequency 
$\Omega $. When the mechanical bath is in a thermal state at temperature $T_{%
{\rm m}}$, the correlation function $\sigma _{mm}^{{\rm in}}\left[ \Omega %
\right] $ of the input field $m^{{\rm in}}$ is equal to\cite
{Callen51,Grassia00} 
\begin{equation}
\sigma _{mm}^{{\rm in}}\left[ \Omega \right] =\frac{1}{2}\coth \frac{\hbar
\left| \Omega \right| }{2k_{B}T_{{\rm m}}},  \label{Eq_Smin}
\end{equation}
where $k_{B}$ is the Boltzman constant.

For a free mechanical oscillator the only remaining force in eq. (\ref{Eq_V}%
) is the input field $m^{{\rm in}}$ and the correlation function of the
mirror velocity $V$ is given by 
\begin{equation}
\left| Z_{{\rm m}}\right| ^{2}\sigma _{VV}\left[ \Omega \right] =\hbar
\left| \Omega \right| H_{{\rm m}}\coth \frac{\hbar \left| \Omega \right| }{%
2k_{B}T_{{\rm m}}}.  \label{Eq_SV}
\end{equation}
Assuming the quality factor $Q$ large compared to $1$, the width $H_{{\rm m}%
}/M$ of the resonance is much smaller than its resonance frequency $\Omega _{%
{\rm m}}$. We can then assume that the spectrum of the fluctuating force
associated with the damping corresponds to a white noise and we replace $%
\Omega $ by $\Omega _{{\rm m}}$ in the right part of this equation. The
velocity spectrum has the usual lorentzian shape corresponding to a
mechanical oscillator in thermal equilibrium at an effective temperature $%
\Theta _{{\rm m}}$ defined as 
\begin{eqnarray}
\left| Z_{{\rm m}}\right| ^{2}\sigma _{VV}\left[ \Omega \right] &=&2H_{{\rm m%
}}k_{B}\Theta _{{\rm m}},  \label{Eq_SV2} \\
k_{B}\Theta _{{\rm m}} &=&\frac{\hbar \Omega _{{\rm m}}}{2}\coth \frac{\hbar
\Omega _{{\rm m}}}{2k_{B}T_{{\rm m}}}.  \label{Eq_Tm}
\end{eqnarray}
At high temperature ($k_{B}T_{{\rm m}}\gg \hbar \Omega _{{\rm m}}/2$) we
find as expected that $\Theta _{{\rm m}}$ is equal to $T_{{\rm m}}$. The
oscillator temperature $\Theta _{{\rm m}}$ however decreases with the bath
temperature $T_{{\rm m}}$ and tends at low temperature towards a limit equal
to $\hbar \Omega _{{\rm m}}/2k_{B}$. This limit is associated with the
zero-point quantum fluctuations of the mechanical oscillator.\ $\Theta _{%
{\rm m}}$ may be written as,

\begin{equation}
k_{B}\Theta _{{\rm m}}=\hbar \Omega _{{\rm m}}\left( n_{\Theta }+\frac{1}{2}%
\right) ,  \label{Eq_ntdef}
\end{equation}
where $n_{\Theta }$ is the number of thermal phonons.\ It is equal to $%
k_{B}T_{{\rm m}}/\hbar \Omega _{{\rm m}}$ at high temperature and reduces to 
$0$ at low temperature. The term $1/2$ in eq. (\ref{Eq_ntdef}) represents
the energy of quantum zero-point fluctuations.

Note that the oscillator temperature $\Theta _{{\rm m}}$ is also related to
the variance $\Delta V^{2}$ of the velocity (equal to the integral of $%
\sigma _{VV}$) by the equipartition theorem, 
\begin{equation}
\frac{1}{2}M\Delta V^{2}=\frac{1}{2}k_{B}\Theta _{{\rm m}}.  \label{Eq_DV2}
\end{equation}

\section{Detection and cold damping}

\label{M_ColdDamping}We now describe the measurement and the feedback loop.
In presence of the intracavity radiation pressure and of an external force,
the mirror velocity can be written from eqs.\ (\ref{Eq_Aa}) and (\ref{Eq_V})
as a function of the input mechanical and optical fields, 
\begin{equation}
Z_{{\rm m}}V=F_{{\rm ext}}-\sqrt{2\hbar \left| \Omega \right| H_{{\rm m}}}m^{%
{\rm in}}+\frac{\sqrt{2\gamma }}{\gamma -i\Omega \tau }\hbar \varkappa
a_{1}^{{\rm in}}.  \label{Eq_V_in}
\end{equation}
This equation clearly shows the two fundamental noise sources for the mirror
motion, corresponding to the coupling to the external bath (second term),
and to the back action of the measurement (last term).

From eqs. (\ref{Eq_A}) ouput fields can be related to input fields and to
mirror velocity,

\begin{mathletters}
\label{Eq_Aout}
\begin{eqnarray}
a_{1}^{{\rm out}} &=&\frac{\gamma +i\Omega \tau }{\gamma -i\Omega \tau }%
a_{1}^{{\rm in}}, \\
a_{2}^{{\rm out}} &=&\frac{\gamma +i\Omega \tau }{\gamma -i\Omega \tau }%
a_{2}^{{\rm in}}+i\frac{2\sqrt{2\gamma }}{\Omega \left( \gamma -i\Omega \tau
\right) }\varkappa V.
\end{eqnarray}
The first equation shows that the reflected amplitude fluctuations are
obtained from the incident ones by a simple phase shift.\ As expected for a
resonant cavity, amplitude fluctuations are not coupled to the mirror
motion. On the contrary the phase of the reflected beam depends on the
cavity length and a measurement of the phase quadrature $a_{2}^{{\rm out}}$
provides information about the mirror motion.\ The result of the measurement
can be described by an estimator $\hat{V}$ of the velocity, which is
proportional to $a_{2}^{{\rm out}}$ and which appears as the sum of $V$ and
of some measurement noise, 
\end{mathletters}
\begin{eqnarray}
\hat{V} &=&-i\frac{\Omega \left( \gamma -i\Omega \tau \right) }{2\sqrt{%
2\gamma }\varkappa }a_{2}^{{\rm out}}  \nonumber \\
&=&V-i\frac{\Omega \left( \gamma +i\Omega \tau \right) }{2\sqrt{2\gamma }%
\varkappa }a_{2}^{{\rm in}}.  \label{Eq_Vest}
\end{eqnarray}
The sensitivity of the velocity measurement is limited by the phase noise of
the incoming beam. The added noise in eq.\ (\ref{Eq_Vest}) is proportional
to $\sqrt{\gamma }$ and inversely proportional to the mean intracavity
amplitude $\alpha _{0}$ (eq. \ref{Eq_K}).\ As a consequence the sensitivity
is increased when cavity finesse or light power are increased.\ One can also
note a frequency filtering by the cavity, the sensitivity being reduced for
frequencies larger than the cavity bandwidth $\Omega _{{\rm cav}}=\gamma
/\tau $.

We apply a feedback force on the mirror proportional to the result of the
measurement, that is to the velocity estimator $\hat{V}$, 
\begin{equation}
F_{{\rm fb}}=-Z_{{\rm fb}}\hat{V},
\end{equation}
where $Z_{{\rm fb}}$ is an impedance which characterizes the transfer
function of the feedback loop. The measurement noise as well as the back
action noise are already present in our analysis. Other noise sources may be
added to the feedback force, such as the quantum fluctuations of the
radiation pressure due to the auxiliary laser beam used for feedback
control, the electronic noise of the feedback loop, or the quantum
efficiency of the detection.\ As it is well known in high sensitivity
measurement, the dominant noise sources are those associated with the first
stage of detection\cite{Courty01}. This result also holds for a quantum
analysis of noise in presence of feedback \cite{Grassia00,Grassia00b}. As a
consequence we neglect in the following these extra noise sources.

The velocity $V_{{\rm fb}}$ in presence of feedback can be deduced from eq.\
(\ref{Eq_V_in}).\ One gets 
\begin{eqnarray}
\left( Z_{{\rm m}}+Z_{{\rm fb}}\right) V_{{\rm fb}} &=&F_{{\rm ext}}-\sqrt{%
2\hbar \left| \Omega \right| H_{{\rm m}}}m^{{\rm in}}+\frac{\sqrt{2\gamma }}{%
\gamma -i\Omega \tau }\hbar \varkappa a_{1}^{{\rm in}}  \nonumber \\
&&+i\frac{\Omega \left( \gamma +i\Omega \tau \right) }{2\sqrt{2\gamma }%
\varkappa }Z_{{\rm fb}}a_{2}^{{\rm in}}.  \label{Eq_Vfb}
\end{eqnarray}
The main effect of feedback is to change the mechanical impedance of the
mirror which becomes the sum of the free mirror impedance and of the
servocontrol impedance, 
\begin{equation}
Z=Z_{{\rm m}}+Z_{{\rm fb}}.  \label{Eq_Z}
\end{equation}
The feedback loop also adds noise to the mirror (last term in eq. \ref
{Eq_Vfb}). It corresponds to a contamination noise introduced by the
feedback mechanism and proportional to the measurement noise of the velocity
estimator. As a consequence there are two different noise sources associated
with light and corresponding to the back action and measurement noises (two
last terms in eq. \ref{Eq_Vfb}).

The general expression of the velocity noise spectrum (without any
assumption on incident field fluctuations) is given by 
\begin{eqnarray}
\left| Z\right| ^{2}\sigma _{VV}^{{\rm fb}} &=&2\hbar \left| \Omega \right|
H_{{\rm m}}\sigma _{{\rm mm}}^{{\rm in}}+\frac{2\gamma }{\left( \gamma
^{2}+\Omega ^{2}\tau ^{2}\right) }\hbar ^{2}\varkappa ^{2}\sigma
_{a_{1}a_{1}}^{{\rm in}}  \nonumber \\
&&+\frac{\Omega ^{2}\left( \gamma ^{2}+\Omega ^{2}\tau ^{2}\right) }{8\gamma
\varkappa ^{2}}\left| Z_{{\rm fb}}\right| ^{2}\sigma _{a_{2}a_{2}}^{{\rm in}}
\nonumber \\
&&-\hbar \Omega 
%TCIMACRO{\func{Im}}%
%BeginExpansion
\mathop{\rm Im}%
%EndExpansion
\left( Z_{{\rm fb}}\right) \sigma _{a_{1}a_{2}}^{{\rm in}}.  \label{Eq_SVfb}
\end{eqnarray}

Let us first examine the effect of feedback when quantum noises can be
neglected as compared to thermal noise. To obtain a cold damping mechanism,
the feedback force $F_{{\rm fb}}$ must correspond to a viscous force, that
is the feedback impedance $Z_{{\rm fb}}$ must be real ($Z_{{\rm fb}}\equiv
H_{{\rm fb}}$, $%
%TCIMACRO{\func{Im}}%
%BeginExpansion
\mathop{\rm Im}%
%EndExpansion
\left( Z_{{\rm fb}}\right) =0$). In this case, the feedback loop changes the
mechanical impedance $Z$ by adding a damping $H_{{\rm fb}}$ to the
mechanical damping $H_{{\rm m}}$. Neglecting the quantum noise of light
(three last terms in eq. \ref{Eq_SVfb}), one finds that the velocity noise
spectrum has the same expression as for a free mechanical oscillator in
thermal equilibrium (eqs.\ \ref{Eq_Smin} and \ref{Eq_SV}), except for the
modification of the mechanical impedance.\ In other words the feedback loop
changes the damping of the mirror without adding any noise. In this
classical analysis of cold damping, the mirror appears to be in a thermal
equilibrium at an effective temperature $\Theta _{{\rm fb}}$ given by, 
\begin{equation}
\Theta _{{\rm fb}}=\frac{H_{{\rm m}}}{H_{{\rm m}}+H_{{\rm fb}}}\Theta _{{\rm %
m}}=\frac{\Theta _{{\rm m}}}{1+g},  \label{Eq_Tfbclassic}
\end{equation}
where the feedback gain $g$ is defined as 
\begin{equation}
g=H_{{\rm fb}}/H_{{\rm m}}.  \label{Eq_g}
\end{equation}
This result shows that the mirror is cooled at an effective temperature
inversely proportional to the gain. It confirms the absence of classical
limits for displacement noise reduction with cold damping\cite{Ritter85}.

Although the main properties of the cold damping mechanism are properly
described here, one gets some inconsistency for large gains since the
effective temperature can decrease down to 0 and it is not limited to the
zero-point effective temperature $\hbar \Omega _{{\rm m}}/2k_{B}$
corresponding to a mechanical oscillator at a null temperature (see eq.\ \ref
{Eq_Tm} and discussion thereafter).\ This is of course due to the fact that
we have neglected all quantum noises. We study in the following how this
result is modified by taking into account these noises.

\section{Noise reduction at resonance}

\label{M_Resonance}We study in this section the maximum noise reduction that
can be obtained at the resonance frequency $\Omega _{{\rm m}}$ when the
measurement and feedback parameters are optimized.\ For this purpose, we
assume that the feedback corresponds to a cold damping mechanism, that is
the feedback impedance $Z_{{\rm fb}}$ is real ($Z_{{\rm fb}}\equiv H_{{\rm fb%
}}$).

The expression of the velocity noise spectrum (eq.\ \ref{Eq_SVfb}) can be
simplified for a mechanical reservoir in a thermal state and for light in a
coherent state.\ In this case the incident field noises are given by eqs. (%
\ref{Eq_Sain}) and (\ref{Eq_Smin}). We assume as in section \ref
{M_OptomechanicalTransducer} that the fluctuating incident force associated
with mechanical damping has a white noise spectrum.\ We furthermore assume
that the cavity bandwidth $\Omega _{{\rm cav}}=\gamma /\tau $ is large
compared to the mechanical resonance frequency $\Omega _{{\rm m}}$ so that
we can neglect the cavity filtering appearing in eq. (\ref{Eq_SVfb}). One
then gets, 
\begin{equation}
\left| Z\right| ^{2}\sigma _{VV}^{{\rm fb}}=2H_{{\rm m}}k_{B}\Theta _{{\rm m}%
}+\frac{2}{\gamma }\hbar ^{2}\varkappa ^{2}+\frac{\Omega ^{2}\gamma }{%
8\varkappa ^{2}}H_{{\rm fb}}^{2}.  \label{Eq_SVfb2}
\end{equation}
The resulting noise spectrum corresponds to the response of the mirror to
the sum of thermal, back action, and measurement noises. The mechanical
impedance $Z$ which characterizes this response takes into account the
presence of feedback.

At resonance, the mechanical impedance takes the simple form, 
\begin{equation}
Z\left[ \Omega _{{\rm m}}\right] =H_{{\rm m}}+H_{{\rm fb}}.  \label{Eq_Zres}
\end{equation}
From eqs. (\ref{Eq_SVfb2}) and (\ref{Eq_Zres}) one can easily show that
arbitrarily small values of the velocity noise at resonance can be reached
by a proper choice of the optomechanical coefficient $\varkappa $ and of the
feedback gain $g=H_{{\rm fb}}/H_{{\rm m}}$. Let us define an optomechanical
parameter $\zeta $ by 
\begin{equation}
\zeta =\frac{4\hbar \varkappa ^{2}}{\gamma \Omega _{{\rm m}}H_{{\rm m}}}.
\label{Eq_Zeta}
\end{equation}
$\zeta $ takes into account the light intensity via $\varkappa $, the cavity
finesse via $\gamma $, and the mechanical response of the mirror. This
parameter can be made much larger than $1$ for a high-finesse cavity, high
incident power, and high mechanical quality factor. $\zeta $ is actually
equal to $4Q\psi _{NL}/\gamma $ where $\psi _{NL}$ is the phase-shift of the
intracavity field due to the mean radiation pressure. $\psi _{NL}$ can be of
the same order as $\gamma $ for realistic experimental parameters\cite
{Heidmann97}.

The velocity noise at resonance is then given by,

\begin{equation}
\sigma _{VV}^{{\rm fb}}\left[ \Omega _{{\rm m}}\right] =\frac{\hbar \Omega _{%
{\rm m}}}{H_{{\rm m}}}\frac{1}{\left( 1+g\right) ^{2}}\left( 2n_{\Theta }+1+%
\frac{\zeta }{2}+\frac{g^{2}}{2\zeta }\right) ,  \label{Eq_SVres}
\end{equation}
where $n_{\Theta }$ is the number of thermal phonons (eq.\ \ref{Eq_ntdef}).\
The velocity noise is now normalized to the noise $\hbar \Omega _{{\rm m}%
}/H_{{\rm m}}$ of the free oscillator at zero temperature (see eqs.\ \ref
{Eq_SV2} and \ref{Eq_ntdef}). Both mechanical noise (term $2n_{\Theta }+1$
in eq.\ \ref{Eq_SVres}) and radiation pressure effects (term $\zeta /2$) are
reduced by the feedback loop.\ In other words, the mirror motion induced by
the quantum fluctuations of radiation pressure is controlled in the same way
as the brownian motion.\ For very large gains $g$, both noises vanish and
the resulting velocity noise is only due to the phase noise added by the
measurement (last term in eq.\ \ref{Eq_SVres}).\ This last noise can also be
made small by increasing the sensitivity of the measurement.

Taking $\zeta \gg 1$, $g\gg \sqrt{\zeta }$, and $g\gg \sqrt{n_{\Theta }}$,
all three terms are small compared to $1$ and the velocity noise of the
cooled mirror becomes smaller than the noise of the free mirror at zero
temperature, 
\begin{equation}
\sigma _{VV}^{{\rm fb}}\left[ \Omega _{{\rm m}}\right] \ll \frac{\hbar
\Omega _{{\rm m}}}{H_{{\rm m}}}.
\end{equation}
As pointed in \cite{Vitali01} a quantum treatment introduces extra noise
sources that were absent in a classical analysis.\ Our results however show
that as in the classical case there is no limit on the noise suppression.

It may appear surprising that the velocity noise at resonance can be made
arbitrarily small. We remind however that cold damping modifies the dynamics
of the oscillator. As a consequence the reduction of velocity noise is made
at the expense of a widening of the mechanical resonance. The result is
fully consistent with quantum mechanics since the velocity noise $\sigma
_{VV}^{{\rm fb}}\left[ \Omega _{{\rm m}}\right] $ is larger than the value $%
\hbar \Omega _{{\rm m}}/\left( H_{{\rm m}}+H_{{\rm fb}}\right) $
corresponding to the zero-point fluctuations of a damped oscillator with the
same mechanical response. In next section we analyze more precisely the
limits due to quantum noise by studying parameters that depend on noise over
all frequencies such as the variance of the velocity or the effective
temperature.

\section{Effective temperature}

\label{M_Temperature}
\begin{figure}
\centerline{\psfig{figure=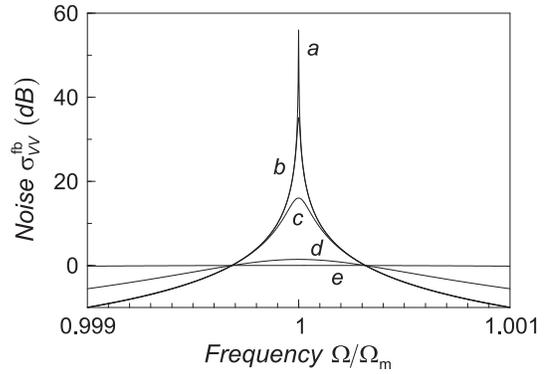,width=7cm}}
\vspace{2mm}
\caption{Velocity noise spectra $\sigma _{VV}^{\rm{fb}}\left[ \Omega \right] $ 
of the mirror in dB scale, without feedback (curve a) and for increasing values of the gain 
$g$ from $10$\ to $10^{4}$ (curves b to e). The spectra have a 
lorentzian shape with an increased width and a reduced amplitude. The limit for high 
gains is related to the quantum noise of light. Parameters are as follows: quality factor 
$Q=10^{6}$, number of thermal phonons $n_{\Theta }=10^{5}$, and optomechanical 
coefficient $\zeta =1$}
\label{Fig_Spectra}
\end{figure}

In presence of feedback the width of the resonance
becomes $\left( 1+g\right) H_{{\rm m}}$. If we assume that this width stays
smaller than the resonance frequency $\Omega _{{\rm m}}$, that is if the
gain $g$ is smaller than the quality factor $Q$, one can neglect the
frequency dependence of the last term in eq.\ (\ref{Eq_SVfb2}) and the
velocity noise spectrum of the cooled mirror at any frequency $\Omega $ can
be written, 
\begin{equation}
\left| Z\right| ^{2}\sigma _{VV}^{{\rm fb}}\left[ \Omega \right] =H_{{\rm m}%
}\hbar \Omega _{{\rm m}}\left( 2n_{\Theta }+1+\frac{\zeta }{2}+\frac{g^{2}}{%
2\zeta }\right) ,  \label{Eq_SVfb3}
\end{equation}
where the number of thermal phonons $n_{\Theta }$ and the optomechanical
parameter $\zeta $ are defined in eqs. (\ref{Eq_ntdef}) and (\ref{Eq_Zeta}).
The resulting noise spectrum is shown in figure \ref{Fig_Spectra} for
different values of the feedback gain $g$. The reduction and widening of the
resonance are clearly visible.

For very large gains, the total impedance $Z$ is proportional to $g$ and the
noise spectrum is limited by the phase noise in the measurement (last term
in eq. \ref{Eq_SVfb3}). In other words, the feedback works in such a way
that its error signal, equal to the velocity estimator $\hat{V}$, goes to $0$%
.\ The mirror velocity $V_{{\rm fb}}$ then reproduces the measurement noise
of the estimator (see eq.\ \ref{Eq_Vest}).

Since the right part in eq.\ (\ref{Eq_SVfb3}) is independent of frequency,
the noise spectrum has always a lorentzian shape. The cooled mirror is thus
equivalent to a harmonic oscillator of resonance frequency $\Omega _{{\rm m}%
} $, damping $\left( 1+g\right) H_{{\rm m}}$, in thermal equilibrium at an
effective temperature $\Theta _{{\rm fb}}$. This temperature can be
determined either from the calculation of the variance $\Delta V^{2}$ as the
integral of the noise spectrum and from the equipartition theorem (eq. \ref
{Eq_DV2}), or by identifying the noise spectrum with the one of a free
oscillator (eq. \ref{Eq_SV2}), taking into account the fact that the damping
is increased by a factor $1+g$.\ One then gets 
\begin{equation}
k_{B}\Theta _{{\rm fb}}=\frac{\hbar \Omega _{{\rm m}}}{2}\frac{1}{1+g}\left(
2n_{\Theta }+1+\frac{\zeta }{2}+\frac{g^{2}}{2\zeta }\right) .
\label{Eq_Tfb}
\end{equation}

\begin{figure}
\centerline{\psfig{figure=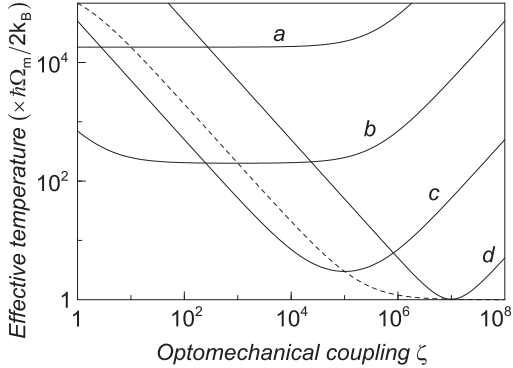,width=6.7cm}}
\vspace{2mm}
\caption{Effective temperature $\Theta _{\rm{fb}}$ of the cooled mirror normalized to 
the zero-temperature limit $\hbar \Omega _{\rm{m}}/2 k_{B}$, as a function of the 
optomechanical coefficient $\zeta $ in logarithmic scales. Curves a to d are obtained 
for gains $g$ equal to $10$, $10^{3}$, $10^{5}$, and $10^{7}$. For each gain, there 
exists a limit which corresponds to a compromise between measurement and 
back action noises. This limit decreases down to $1$ for large gains as the residual
thermal noise is reduced (dashed curve). The 
initial number of thermal phonons $n_{\Theta }$ is equal to $10^{5}$}
\label{Fig_Temp}
\end{figure}

The effective temperature $\Theta _{{\rm fb}}$ of the cooled mirror is
plotted in figure \ref{Fig_Temp} as a function of the optomechanical
coefficient $\zeta $, for different values of the gain $g$ and for an
initial number of thermal phonons $n_{\Theta }$ equal to $10^{5}$. Efficient
reduction of temperature can be achieved as soon as $g$ is larger than $%
n_{\Theta }$ (curves {\it c} and {\it d}). The effective temperature is then
determined by quantum noise.\ As in usual optical measurements\cite
{Caves81,Braginsky92}, phase noise of the measurement is dominant for low
values of $\zeta $ (term $g^{2}/2\zeta $ in eq.\ \ref{Eq_Tfb}), and back
action of radiation pressure is dominant for high $\zeta $ (term $\zeta /2$
in eq.\ \ref{Eq_Tfb}). A quantum limit is reached for a precise value of $%
\zeta $ which corresponds to the case where both noises are equal, 
\begin{equation}
\zeta ^{opt}=g.
\end{equation}
The minimum effective temperature is then, 
\begin{equation}
k_{B}\Theta _{{\rm fb}}^{opt}=\hbar \Omega _{{\rm m}}\left( \frac{n_{\Theta }%
}{1+g}+\frac{1}{2}\right) .  \label{Eq_Topt}
\end{equation}
We have thus found that for a given value of $g$, there exists a limit for
the effective temperature of the mirror.\ This limit evolves from the
initial temperature $\Theta _{{\rm m}}$ for no gain, down to $\hbar \Omega _{%
{\rm m}}/2k_{B}$ for an infinite gain (dashed curve in figure \ref{Fig_Temp}%
).\ This minimum corresponds to the effective temperature of a free harmonic
oscillator coupled to a thermal bath at zero temperature (eq.\ \ref{Eq_Tm}).

Comparison between the cases of a cold damped mirror (eq. \ref{Eq_Topt}) and
a free mirror (eq. \ref{Eq_ntdef}) shows that the number of thermal phonons
is reduced by a factor $1+g$. This is the same reduction than with classical
cold damping (eq. \ref{Eq_Tfbclassic}).\ On the contrary the quantum part
(term $1/2$) remains unchanged. One can then define an effective number $%
n_{\Theta {\rm fb}}$ of thermal phonons in presence of feedback as, 
\begin{equation}
k_{B}\Theta _{{\rm fb}}=\hbar \Omega _{{\rm m}}\left( n_{\Theta {\rm fb}}+%
\frac{1}{2}\right) .
\end{equation}
The minimum number of thermal phonons is reached for $\zeta =g$ and is
inversely proportional to the gain, 
\begin{equation}
n_{\Theta {\rm fb}}^{opt}=\frac{n_{\Theta }}{1+g}.
\end{equation}
This equation is the quantum generalization of the classical behavior of
cold damping (eq. \ref{Eq_Tfbclassic}).

\section{Quantum limits of cold damping}

\label{M_QuantumLimit}We revisit now the results obtained in previous
sections within the framework of a general analysis of the noise added by a
feedback mechanism. We will show that a minimum noise is imposed by quantum
mechanics and that the cold damped mirror reaches this limit.

When a quantum system is coupled to a thermal bath, the quantum
fluctuations-dissipation theorem provides a relation between the commutator
of the force and the noise added by the coupling. A mean to obtain this
relation is to analyze the unitarity of the input-output relations of the
system. This kind of analysis also holds for active systems and provides
lower bounds to the noise associated with amplifying devices\cite
{Haus62,Caves82,Courty99}.

This kind of analysis may be used to study the noise associated to a linear
feedback. Let us consider a system characterized by a velocity $V$ and an
impedance $Z_{{\rm m}}$ which is linearly fed back by a force $F_{{\rm fb}}$
applied with an impedance $Z_{{\rm fb}}$.\ Equations of the system can be
written as the general relations, 
\begin{mathletters}
\label{Eq_Feed}
\begin{eqnarray}
Z_{{\rm m}}V &=&F_{{\rm ext}}-\sqrt{2\hbar \left| \Omega \right| H_{{\rm m}}}%
m^{{\rm in}}+F_{{\rm fb}}, \\
F_{{\rm fb}} &=&-Z_{{\rm fb}}V+F_{{\rm fb}}^{{\rm in}},  \label{Eq_Feed_b}
\end{eqnarray}
where $m^{{\rm in}}$ is the input field associated with the free system.\ To
preserve the commutation relations one has to introduce a force $F_{{\rm fb}%
}^{{\rm in}}$ in eq.\ (\ref{Eq_Feed_b}) which appears as a noise term for
the feedback force.\ To write the input-output transformation one has also
to introduce an output field $m^{{\rm out}}$ defined in the same way as the
output light field $a^{{\rm out}}$ for an optical system\cite{Grassia00}, 
\end{mathletters}
\begin{equation}
m^{{\rm out}}=m^{{\rm in}}+\sqrt{\frac{2H_{{\rm m}}}{\hbar \left| \Omega
\right| }}V.
\end{equation}
The output field $m^{{\rm out}}$ is a free field which can be related to the
incident fields $m^{{\rm in}}$ and $F_{{\rm fb}}^{{\rm in}}$ from eqs. (\ref
{Eq_Feed}).\ One gets, 
\begin{equation}
m^{{\rm out}}=\frac{Z-2H_{{\rm m}}}{Z}m^{{\rm in}}+\sqrt{\frac{2H_{{\rm m}}}{%
\hbar \left| \Omega \right| }}\frac{1}{Z}F_{{\rm fb}}^{{\rm in}},
\label{Eq_mout}
\end{equation}
where $Z$ is the impedance in presence of feedback (eq. \ref{Eq_Z}).

The unitarity of input-output transformations implies that commutators of
the output field $m^{{\rm out}}$ and of the input field $m^{{\rm in}}$ are
identical (eq. \ref{Eq_Commut}). As a consequence one gets from eq.\ (\ref
{Eq_mout}) the commutator of the noise added by feedback, 
\begin{equation}
\left[ F_{{\rm fb}}^{{\rm in}}\left[ \Omega \right] ,F_{{\rm fb}}^{{\rm in}}%
\left[ \Omega ^{\prime }\right] \right] =2\pi \delta \left( \Omega +\Omega
^{\prime }\right) 2\hbar \Omega H_{{\rm fb}}.  \label{Eq_FfbCommut}
\end{equation}
This commutator implies a Heisenberg inequality on the correlation function $%
\sigma _{F_{{\rm fb}}F_{{\rm fb}}}^{{\rm in}}$ of the force $F_{{\rm fb}}^{%
{\rm in}}$, 
\begin{equation}
\sigma _{F_{{\rm fb}}F_{{\rm fb}}}^{{\rm in}}\left[ \Omega \right] \geq
\hbar \left| \Omega \right| H_{{\rm fb}}.  \label{Eq_FfbCorrel}
\end{equation}

This feedback noise has of course important consequences for the resulting
noise spectrum of the system.\ The correlation function $\sigma _{VV}^{{\rm %
fb}}$ in presence of feedback can be calculated from eqs. (\ref{Eq_Feed})
and one gets, 
\begin{equation}
\left| Z\right| ^{2}\sigma _{VV}^{{\rm fb}}=2\hbar \left| \Omega \right| H_{%
{\rm m}}\sigma _{mm}^{{\rm in}}+\sigma _{F_{{\rm fb}}F_{{\rm fb}}}^{{\rm in}%
}.  \label{Eq_SVfb4}
\end{equation}
As in previous sections we assume that the total impedance $Z$ is
characterized by a width of the resonance much smaller than its resonance
frequency $\Omega _{{\rm m}}$. We can then replace $\Omega $ by $\Omega _{%
{\rm m}}$ in the right part of eq. (\ref{Eq_SVfb4}). The servocontrolled
system is found to have the same velocity noise spectrum than an oscillator
with an impedance $Z$, in thermal equilibrium at an effective temperature $%
\Theta _{{\rm fb}}$ given by 
\begin{equation}
\Theta _{{\rm fb}}=\frac{H_{{\rm m}}\Theta _{{\rm m}}+H_{{\rm fb}}\Theta _{%
{\rm fb}}^{{\rm in}}}{H_{{\rm m}}+H_{{\rm fb}}},  \label{Eq_TFeed}
\end{equation}
where $\Theta _{{\rm fb}}^{{\rm in}}$ is the effective noise temperature of
the feedback force defined as, 
\begin{equation}
\sigma _{F_{{\rm fb}}F_{{\rm fb}}}^{{\rm in}}\left[ \Omega _{{\rm m}}\right]
=2H_{{\rm fb}}k_{B}\Theta _{{\rm fb}}^{{\rm in}}.
\end{equation}
The effective temperature of the servocontrolled system is the average of
the temperatures of the free system and of the feedback noise, weighted by
the corresponding damping coefficients.

If the feedback noise corresponds to a coupling with a thermal bath at the
initial temperature ($\Theta _{{\rm fb}}^{{\rm in}}=\Theta _{{\rm m}}$), one
finds that the feedback loop does not change the temperature.\ The
servocontrol modifies the impedance of the system without any cooling
effect. In this case it is equivalent to a modification of the impedance by
passive means.

Quantum mechanics does not however prevent to use a feedback loop which has
a lower noise temperature.\ With active elements or in presence of frequency
transfer\cite{Grassia00}, the noise temperature is not the physical
temperature of the device but it is determined by the physical process
coming into play. The effective temperature of the servocontrolled system
can therefore be reduced, down to a limit imposed on feedback noise by the
Heisenberg inequality (eq. \ref{Eq_FfbCorrel}), 
\begin{equation}
k_{B}\Theta _{{\rm fb}}^{{\rm in}}\geq \frac{\hbar \Omega _{{\rm m}}}{2}.
\label{Eq_Heisenberg}
\end{equation}
In this situation, the whole detection and feedback system acts on the
oscillator as a coupling with a thermal reservoir at an effective
temperature $\Theta _{{\rm fb}}^{{\rm in}}$ lower than $\Theta _{{\rm m}}$.

We analyze now the performance of the optomechanical cold damping in light
of these limits. The added force $F_{{\rm fb}}^{{\rm in}}$ can be deduced
from eq.\ (\ref{Eq_Vfb}), 
\begin{equation}
F_{{\rm fb}}^{{\rm in}}=\frac{\sqrt{2\gamma }}{\gamma -i\Omega \tau }\hbar
\varkappa a_{{\rm 1}}^{{\rm in}}+i\frac{\Omega \left( \gamma +i\Omega \tau
\right) }{2\sqrt{2\gamma }\varkappa }Z_{{\rm fb}}a_{{\rm 2}}^{{\rm in}}.
\label{Eq_Ffbin}
\end{equation}
It can be checked that this force verifies the required commutation relation
(eq.\ \ref{Eq_FfbCommut}). From eqs.\ (\ref{Eq_V_in}) and (\ref{Eq_Aout}),
it is even possible to write input-output relations for all the fields
(mechanical field $m$, light fields $a_{1}$ and $a_{2}$) and to verify the
global unitarity of the whole transformation.

Let us first examine the case of a dissipative feedback corresponding to the
cold damping situation ($%
%TCIMACRO{\func{Im}}%
%BeginExpansion
\mathop{\rm Im}%
%EndExpansion
\left( Z_{{\rm fb}}\right) =0$, coherent incident field, and cavity
bandwidth much larger than the mechanical resonance frequency). The noise
spectrum of the added force can be derived from eq. (\ref{Eq_Ffbin}) and
leads to the feedback noise temperature, 
\begin{equation}
k_{B}\Theta _{{\rm fb}}^{{\rm in}}=\frac{\hbar \Omega _{{\rm m}}}{2}\left( 
\frac{\zeta }{2g}+\frac{g}{2\zeta }\right) ,
\end{equation}
where the feedback gain $g$ and the optomechanical parameter $\zeta $ have
been defined in eqs. (\ref{Eq_g}) and (\ref{Eq_Zeta}).\ The feedback noise
temperature obviously satisfies the Heisenberg inequality (eq.\ \ref
{Eq_Heisenberg}) and reaches its minimum value $\hbar \Omega _{{\rm m}}/2$
when $\zeta $ is equal to $g$. This equation sheds a new light on the
results of previous section.\ The effect of cold damping can be interpreted
in this case as a coupling with a thermal reservoir at zero temperature.

Let us finally examine a more general situation for which the incident field
can be in a squeezed state and the feedback impedance can have a non zero
reactive part.\ In this case, the added noise derived from eq. (\ref
{Eq_Ffbin}) leads to a feedback noise temperature, 
\begin{eqnarray}
k_{B}\Theta _{{\rm fb}}^{{\rm in}} &=&\frac{\hbar \Omega _{{\rm m}}}{2}\left[
\frac{\left| Z_{{\rm fb}}\right| }{H_{{\rm fb}}}\left( \frac{\zeta }{2g}%
\sigma _{a_{1}a_{1}}^{{\rm in}}+\frac{g}{2\zeta }\sigma _{a_{2}a_{2}}^{{\rm %
in}}\right) \right.  \nonumber \\
&&\left. -\frac{%
%TCIMACRO{\func{Im}}%
%BeginExpansion
\mathop{\rm Im}%
%EndExpansion
\left( Z_{{\rm fb}}\right) }{H_{{\rm fb}}}\sigma _{a_{1}a_{2}}^{{\rm in}}%
\right] ,
\end{eqnarray}
where the gain $g$ is now defined as $\left| Z_{{\rm fb}}\right| /H_{{\rm m}%
} $.

With light fluctuations corresponding to a coherent state, the
intensity-phase correlations $\sigma _{a_{1}a_{2}}^{{\rm in}}$ are zero and
the minimal temperature, reached for $\zeta =g$, is equal to 
\begin{equation}
k_{B}\Theta _{{\rm fb}}^{{\rm in}}=\frac{\hbar \Omega _{{\rm m}}}{2}\frac{%
\left| Z_{{\rm fb}}\right| }{H_{{\rm fb}}}.
\end{equation}
If the reactive part of the feedback impedance is not zero, the effective
temperature is larger than the minimum imposed by the Heisenberg inequality
(eq. \ref{Eq_Heisenberg}).

It is however possible to reach the optimum value $\hbar \Omega _{{\rm m}}/2$
with an incident squeezed state, where intensity and phase noises are
correlated.\ Correlation functions of the incident field must be equal to 
\begin{mathletters}
\label{Eq_Squeezing}
\begin{eqnarray}
\sigma _{a_{1}a_{1}}^{{\rm in}} &=&\frac{g}{\zeta }\frac{\left| Z_{{\rm fb}%
}\right| }{H_{{\rm fb}}}, \\
\sigma _{a_{2}a_{2}}^{{\rm in}} &=&\frac{\zeta }{g}\frac{\left| Z_{{\rm fb}%
}\right| }{H_{{\rm fb}}}, \\
\sigma _{a_{1}a_{2}}^{{\rm in}} &=&\frac{%
%TCIMACRO{\func{Im}}%
%BeginExpansion
\mathop{\rm Im}%
%EndExpansion
\left( Z_{{\rm fb}}\right) }{H_{{\rm fb}}}.
\end{eqnarray}
Note that this state is a minimum state for the generalized Heisenberg
inequality of the field\cite{Reynaud92}, 
\end{mathletters}
\begin{equation}
\sigma _{a_{1}a_{1}}^{{\rm in}}\sigma _{a_{2}a_{2}}^{{\rm in}}-\left( \sigma
_{a_{1}a_{2}}^{{\rm in}}\right) ^{2}=1.
\end{equation}

Two situations are of particular interest.\ First, we still consider a cold
damping mechanism ($%
%TCIMACRO{\func{Im}}%
%BeginExpansion
\mathop{\rm Im}%
%EndExpansion
\left( Z_{{\rm fb}}\right) =0$), but with an incident field which phase is
squeezed by a factor $e^{-\xi }$, 
\begin{equation}
\sigma _{a_{1}a_{1}}^{{\rm in}}=e^{\xi },\ \sigma _{a_{2}a_{2}}^{{\rm in}%
}=e^{-\xi },\ \sigma _{a_{1}a_{2}}^{{\rm in}}=0.
\end{equation}
In this case, the zero noise temperature can be reached for a smaller
optomechanical parameter $\zeta $, equal to $e^{-\xi }g$. The effect is
actually similar to the one that can be obtained in interferometric
measurements\cite{Caves81}.\ The incident phase-squeezed state reduces the
phase noise of the measurement, at the expense of an increase of the back
action noise due to radiation pressure.\ As a consequence the standard
quantum limit can be reached for a lower light power. For the cold damped
mirror, this corresponds to a translation towards the left of the curves in
figure \ref{Fig_Temp}.

In the second situation we consider that the reactive part of the feedback
impedance is not zero. This may be wanted to change both the mechanical
damping and the mechanical resonance position. A squeezed state with
intensity-phase correlations is required to reach the minimum of added
noise. For a value of the optomechanical parameter $\zeta $ equal to $g$,
the squeezed quadrature has to be rotated by an angle of $45%
%TCIMACRO{\UNICODE[m]{0xb0}}%
%BeginExpansion
{{}^\circ}%
%EndExpansion
$ with respect to phase and intensity quadratures, and the squeezing factor
must be equal to 
\begin{equation}
e^{-\xi }=\frac{\left| Z_{{\rm fb}}\right| -\left| 
%TCIMACRO{\func{Im}}%
%BeginExpansion
\mathop{\rm Im}%
%EndExpansion
\left( Z_{{\rm fb}}\right) \right| }{H_{{\rm fb}}}.
\end{equation}
The optimum noise is then reached for a finite squeezing factor, except for
a purely reactive feedback ($H_{{\rm fb}}=0$).\ In this case, the zero noise
temperature is only a limit corresponding to an infinite squeezing.

The zero noise temperature can also be reached for arbitrary values of the
optomechanical parameter $\zeta $ ($\zeta \neq g$), by a proper choice of
the squeezed quadrature and of the squeezing factor (see eqs.\ \ref
{Eq_Squeezing}).\ It is in particular possible to reduce the optomechanical
parameter at the expense of an increase of the squeezing factor.

\section{Conclusion}

The whole detection and feedback system used in cold damping techniques
allows to simulate a thermal reservoir at zero temperature. It is then
possible in principle to cool the oscillator down to its zero-point quantum
fluctuations.

The optomechanical system is well adapted to cold damping and can be
optimized to reach the limits imposed by quantum mechanics. In this system,
the performance limits are due to the Heisenberg inequality on the intensity
and phase of the detection beam. They correspond to the general limits of
cold damping.

As it is the case for classical cold damping, thermal fluctuations are
reduced by a factor inversely proportional to the feedback gain.\ Zero-point
quantum fluctuations of the oscillator are however left unchanged by
feedback and provide a limit to the reduction of the oscillator energy.

Reduction of the effective temperature is accompanied by an increase of the
effective damping of the mirror.\ Although the energy is limited by
zero-point fluctuations, we have shown that arbitrarily large noise
reduction can be achieved at a given frequency.

One may finally wonder how these limits can be experimentally observed.\ The
residual brownian motion of the cold damped mirror can be measured by a
second displacement sensor, or equivalently by using a second independent
light beam in the high-finesse cavity.\ One has however to take into account
the quantum noises associated with this second beam.\ As for the first
intracavity beam, one finds that radiation pressure effects of the second
beam are controlled by the feedback loop so that the sensitivity of the
measurement is only limited by the phase noise, which can be made
arbitrarily small by increasing the light power.

\end{document}